\documentstyle[11pt]{article}

\normalsize

 \def\lagr{\hbox{$\cal L$}} 
\def\ham{\hbox{$\cal H$}}  \def\dalam{\hbox 
{\vrule\vbox{\hrule\hbox to 1ex{ \hfill}\kern 1 
ex\hrule}\vrule}} \def\1/2{\hbox{$ {1 \over 2}$ }} \def\i/h{{i 
\over \h}} \def\<{\langle} \def\>{\rangle} \def\th{\tanh} 
\def\ch{\cosh} \def\sh{\sinh} \def\sign{\hbox{sign }} \def\h 
{\hbar} \def\v{\vec} \def\pd{\partial} \def\({\left(} 
\def\[{\left[} \def\){\right)} \def\]{\right]}

\def\a{\alpha}
\def\b{\beta}
  \def\G{\Gamma}
\def\d{\delta}  
   \def\L{\Lambda}

\def\x{\xi}
\def\c{\chi}
\def\m{\mu}
\def\n{\nu}
\def\t{\tau}
\def\p{\pi}

\def\w{\omega}
\def\tt{\theta}

\def\f{\varphi} \def\F{\Phi}
\def\e{\varepsilon}

\begin{document}
\bf

\begin{center}
UV-regularization of field discontinuities
\end{center}

\begin{center}
K.Sveshnikov, P.Silaev, and I.Cherednikov 
\end{center}

\it
\begin{center}
Department of Physics and Institute of Theoretical Microphysics,
\break Moscow State University,  Moscow 119899, Russia
\end{center}
\vskip 0.3 true cm

\rm
\leftskip 1 true cm
\rightskip 1 true cm
\small \baselineskip 10 pt

A nonperturbative  regularization of UV-divergencies, caused by 
finite discontinuities in the field configuration, is discussed 
in the context of 1+1-dimensional kink models. The relationship 
between this procedure and the appearance of "quantum copies" of 
classical kink solutions is studied in detail and confirmed by 
conventional methods of soliton quantization. 

\vskip 0.5 true cm

\leftskip 0 true cm
\rightskip 0 true cm
\normalsize \baselineskip 13pt

Particlelike classical solutions (henceforth kinks) and their 
quantum descendants play a significant role in description of 
extended objects in the QFT [1-4].  Let $\f_c(x)$ be the 
stationary solution of such type, then, as a rule, it 
interpolates smoothly between static field vacua $\f_{vac}$, 
corresponding to the global minima of the (effective) energy 
functional.  Now let us note, that with each smooth solution 
$\f_c(x)$ one can connect a discontinuous steplike configuration
$\f_d(x)$ with the same topology, whose main difference from  
$\f_c(x)$ is the vanishing size of the transition region, where $\f 
\not = \f_{vac}$. For instance, in 1+1-dimensions the familiar kink 
solution is $\f_c(x)=\th x$, whose discontinuous analog is nothing 
else, but $\f_d(x)= \sign x$. The same situation takes place for 
more spatial dimensions too, e.g., for skyrmeon-like configurations.  
Namely, the discontinuous analog for the hedgehog $U(\v r)=\exp [i\v 
n \v \t F(r) ]$ can be easily constructed via  replacing the radial 
chiral angle function $F(r)$, which varies continuously from 
$F(0)=\p n$ to $F(\infty)=0$, by $F_d(r)=\p n \ \tt(R-r)$ with some 
finite $R$.

So far, such discontinuous configurations have been dismissed as 
unimportant physically, on account of two main reasons. The first one is 
that, purely classically, they do not obey the field equations of 
motion in the points of discontinuity, where the 
derivatives of jumps will cause the appearance of singular sources 
in the r.h.s.  The second reason is that  at first glance their 
energy should be infinite due to singularities in the kinetic term.  
However, in fact the situation turns out to be more subtle, since in 
the quantum theory certain almost discontinuous configurations of 
such type appear (at least formally) as proper solutions of field 
equations in the framework of covariant collective variables [4].  
It is worth-while noticing, that in this framework the corresponding 
modification of field equations originates only from the special 
operator-valued change of field variables in the model without any 
changes in the initial Lagrangean. At the same time, the main reason 
for diverging energy of such discontinuous configurations with 
finite jumps is that in this case the high-frequency component of 
the field $\tilde \f (k)$ decreases to slow for $|k| \to \infty$, 
and so the abovementioned divergencies are of an apparent UV-type.  
However, we know, that at least for some QFT models there exists a 
reliable way to remove the UV-divergencies via regularization 
followed by corresponding subtractions. So there appears a natural 
assumption, that there might exist a reasonable subtracting 
procedure, which removes the singularities, caused by such finite 
jumps.  It should be noted, however, that such subtracting should be 
quite different from the standard renormalization technique. The 
latter removes the infinities only for perturbations around the 
ground state, and reveals actually a similar structure in the vacuum 
as well as in sectors with smooth classical background fields like 
kinks or solitons [1], whereas in the present case one assumes a 
nonperturbative procedure, which renormalizes properly the jumps of 
the field and the divergency in their energy, what obviously cannot 
be obtained in any finite order of the perturbative expansion. 

The purpose of this letter is to present arguments in favour of 
existence of such nonperturbative subtracting  procedure within 
the context of 1+1-dimensional kink theory, described by the 
Lagrangean $$ \lagr (\f)=\1/2 (\pd_{\m} \f)^2 -V(\f) \ , 
\eqno(1)$$ which possesses a stationary kink solution $\f_c(x)$. 
Moreover, we'll see, that after such procedure 
there appears a kind of quantum smoothing of  such finite jumps 
in the sense, that we'll be able to find a series of specific smooth 
c-number solutions of original field equations, which are of the same 
topology as the classical kink, but nonanalytically depend 
on the (effective) Planck constant $\h$, i.e., on  the factor, that enters 
the r.h.s. of the CCR $[\f (x,t), \p (y,t) ]=i\h \d(x-y)$, and possess the 
corresponding discontinuous configuration $\f_d(x)$ as the improper limit 
for $\h \to 0$. 

In implementing this program the crucial point  is to 
quantize the kinks covariantly by means of the relativistic 
(Newton-Wigner) c.m.s.  coordinate [5], which in 1+1-dimensions is given 
by $$ q= {1\over 2} \({1 \over H}L+L{1 \over H}\), \eqno(2)$$ where $H$, 
$L$ and $P$ are the generators of 1+1-dimensional Poincare algebra $$ 
[L,H] =i\h P,\quad [L,P] =i\h H, \quad [H,P]=0.  \eqno(3) $$ The physical 
meaning of the operator $q$ implies that it should enter the field 
arguments only in the combination $(x-q(t))$ .  It would be convenient to 
use the following substitution [6-8] $$x \to \x(x,t), \eqno(4)$$ where $$ 
\x(x,t)=Hx-Pt-L=H^{1/2}(x-q(t))H^{1/2}.  \eqno(5)$$ The 
dimensionless coordinate  $\x(x,t)$, defined in such a way, 
behaves under the Poincare group transformations  $x^{\m}\to 
x'^{\m}=\L^{\m}_{\n} x^{\n}+a^{\m}$ like a covariant scalar, 
i.e.  $$ U^+(\L,a)\x(x)U(\L,a)=\x [\L^{-1}(x-a)], \eqno(6)$$ 
provided the generators of unitary transformations $U(\L,a)$ are 
given by the algebra (3).  Through this property any c-number 
function of it will be a covariant scalar as well.  So we are 
led to represent the Heisenberg field in the kink sector 
(assuming the validity of weak-coupling expansion in $\sqrt \h$ 
for mesonic fluctuations) as follows 
$$\f(x,t)=f(\x(x,t))+\F(\xi(x,t),t).  \eqno(7)$$ In eq.(7)  
$f(\x(x,t))$ describes the quantum kink, provided $f(\x)$ is a 
c-number kink shape function, while $\F(\xi(x,t),t)$  denotes 
the mesonic modes. The latter, considered as a function of the 
c-number arguments, commutes with the generator of spatial 
translations $P$, while the commutation relations between 
$\F(\xi,t)$ and other generators of the Poincare group are 
determined from the covariance properties of the whole field 
$\f(x,t)$ [6-8].  

It should be stressed, that eqs.(2-7) are nothing else, but a 
special change of field variables, which restores the space-time 
symmetry, broken by the nontrivial c-number solution,  directly on 
the quantum level. The number of degrees of freedom is preserved by 
equating the Poincare algebra (3) to the corresponding Noether 
expressions for space-time generators, derived from the initial 
field Lagrangean (1) [4,8].  Moreover, one can show quite 
unambigously, that it is indeed the repr.(7), that  appears as the 
final result of collecting together all the secular terms in the 
straightforward perturbation expansion around the stationary 
classical kink solution [9]. Within expansion in the inverse powers 
of the kink mass this representation turns out to be equivalent to 
the canonical methods of soliton quantization, exploring  the 
collective coordinate $X(t)$ [10-12], as well as to quasiclassical 
expansions, that appear in the BRST-approach [13-15]. 

At the same time, the appearance of the NW coordinate (2) in field 
arguments gives rise for some effects of an apparently 
nonperturbative nature, which take place directly for the kink 
solutions of the quantized theory.  The main circumstance here is, 
that within the repr.(7) the dynamical eq.  of motion for the 
c-number kink shape $f(\x)$ acquires a very special form, where all 
the differential operators are replaced by finite differences with 
the step $i\h$, while the singularities, caused by differentiating 
of jumps, show up as specific pole terms.  These pole terms can be 
represented in the form, quite analogous to the conventional 
UV-divergencies of the field, calculated in the rapidity 
representation. Postulating the subtracting 
of such pole terms, we'll get a consistent scheme of covariant kink 
quantization, where the finite jumps of the field are 
UV-renormalized  and instead of divergencies give rise for a 
series of smooth solutions of field equations, which reproduce 
the original jump as the improper limit for $\h \to 0$, but  
their energy remains finite and vanishes, when the size of 
the solution tends to zero.

To start with, let us  insert the repr.(7) into the initial 
field eq. of motion $$ \dalam \f + V'(\f)=0, \eqno(8)$$ with 
disregard to the mesonic part. Due to the explicit covariance of the 
kink field $f(\xi(x,t))$ the calculation of space-time derivatives 
reduces to evaluation of commutators with translation generators 
$$ \pd_{\m}f(\x(x,t)) = {i \over \h } \[ P_{\m}, f(\x(x,t)) \].  
\eqno(9)$$ The calculation of such commutators proceeds as 
follows.  By means of the light-cone operators $$ K_{\pm} =H\pm 
P  \eqno(10)$$ one obtains for any integer power of $\x$ $$ 
K_{\pm} \xi^n (x,t)= [ \xi(x,t) \pm i\h ] ^n K_{\pm}, 
\eqno(11)$$ whence $$ \dalam _{xt} \xi^n = \left( {i \over \h } 
\right)^2 [K_+, [K_-,\xi^n]] = {M^2 \over \h^2} [ 
(\xi+i\h)^n+(\xi-i\h)^n-2\xi^n], \eqno(12)$$ where 
$M^2=K_+K_-=H^2-P^2$ is the mass operator squared, which 
commutes with the algebra (3) and so with any $f(\xi(x,t))$.  
Therefore for any entire function $f(\x)$ there holds $$ \dalam 
_{xt} f(\xi(x,t)) =\sum \limits _n f_n \ \dalam _{xt} \xi^n 
(x,t) = {M^2 \over \h^2} [ f(\xi+i\h)+f(\xi-i\h)-2f(\xi)](x,t), 
\eqno(13)$$ whence we get the following eq. for the kink shape 
$$ f(\x +i \h)+f(\x -i\h) -2f(\x)+{\h^2 \over M^2} V'(f(\x))=0.  
\eqno(14)$$ In eq.(14) the argument $\x$ is already a c-number, 
which varies for $- \infty$ to $+ \infty$,  since the spectrum 
of the operator $\x(x,t)$ coincides with that of $(-L)$ and 
fills in the real axis. 

However, the situation changes drastically in the case of 
 $f(\x)$ being meromorphic.  Recalling the Mittag-Leffler 
 theorem [16], we can represent $f(z)$ as a series of  its 
principal parts, which converges in each regular point of $f(z)$ 
in the complex plane, $$ f(z)= \sum_n \sum \limits_{k=1} ^{m_n} 
{ c^{(n)}_{-k} \over (z-a_n)^k } + \sum \limits _{k=1} ^{m} c_k 
z^k, \eqno(15)$$  thereby it is assumed, that the poles $a_n$ 
don't lye on the lines $Im \ z= 0, \pm \h$. At first glance, the 
commutation relations between $K_{\pm}$ and inverse powers $(\xi 
-a_n)^{-k}$ should automatically follow from eq.(11). However, 
this statement is incorrect, since first of all one has to 
define selfconsistently what should be meant under $(\xi 
-a_n)^{-k}$. Namely, the natural definition of any function of 
the operator $\x$ should be made via corresponding spectral 
expansion into the complete set of eigenvectors of $\x$.  Such 
eigenvectors can be easily found by noticing, that the mass 
operator $M$ becomes a good quantum number, namely the kink 
mass, when the mesons are neglected. In this case we can employ 
the rapidity representation, where the Lorentz boost $L$ takes 
the form $$ L=i \h {\pd \over \pd \c} \ , \eqno(16)$$ the 
canonically conjugated to $L$ variable $\c$ is the rapidity, 
while the space-time translation generators are $H=M\ch \c$, 
$P=M\sh \c$.  The eigenvectors of $L$ and $\x=\xi(0,0)=-L$ are 
given by $$ \< \c | L \>= (2 \pi \h)^{-1/2} \exp \left( -{i 
\over \h} L \chi \right) \ , \ \langle \chi | \xi \rangle = (2 
 \pi \h)^{-1/2} \exp \left( {i \over \h} \xi \chi \right).  
\eqno(17) $$ By means of eqs. (16) and (17) one can easily 
verify, that there hold the following commutation relations 
between $K_{\pm}$ and $(\xi-a_n)^{-1}$ $$ K_{\pm} {1 \over \xi 
-a}={1 \over \xi \pm i\h -a} K_{\pm} \pm 2\pi i K_{\pm} \d(\x-a) 
\ , \eqno(18)$$ where the term with $\d$-function appears for 
$K_+ $, when $0< Im \ a <\h$, and for $K_- $, when $0>Im \ a > 
-\h$, while the $\d$-function of the complex operator argument 
is defined through its spectral transform as well. Using eq.(18) 
and the Mittag-Leffler theorem, we find for any meromorphic 
$f(\x)$ $$ K_{\pm} f(\x)=f(\x \pm i\h) K_{\pm} \pm 2\pi i 
K_{\pm} \sum \G_n \d(\x-a_n) \ , \eqno(19)$$ where $\G_n$ are 
the residues of $f(\x)$ in the poles $a_n$, and to the term with 
$\d$-functions contribute only those poles, which lye in the 
strips $0< Im \ a_n <\h$ for $K_+ $, and $0>Im \ a_n >-\h$ for 
$K_- $.

From the eq.(19) we immediately find, that whenever the function 
$f(z)$ possesses the poles in the strip  $|Im \ z |<\h$, the 
differentiation of the operator $f(\x(x,t))$ yields besides the 
regular differences like in eq.(13) the additional singular pole 
terms, containing  $\d(\x-a_n)$. This are indeed these pole 
terms, which cause the divergence in the kinetic energy, since 
the latter is at least quadratic in derivatives and so will 
contain the integral over $d\x$ with $\d$-functions squared.  To 
make this relationship between poles in the strip $|Im \ z |<\h$ 
and divergencies in the energy more transparent, let us achieve 
the  same conclusion by considering the matrix elements of the 
kink operator within the relativized version of the 
Goldstone-Jackiw approach to soliton quantization [17], when the 
kink states are taken as the rapidity eigenstates  $|\c \>$ with 
the energy-momentum $\e=M\ch \c$, $p=M\sh \c$.

Following GJ, we define the kink formfactor as $$ \tilde f(\c 
-\c')=\< \c | \f(0,0) |\c' \>=\int \!  d\xi \ \< \chi|\xi \> \ 
f(\xi) \ \<\xi|\chi'\> =(2\pi \h)^{-1} \int \!  d\xi \ f(\xi) 
\exp \[ {i \over \h} \xi (\chi-\chi') \].  \eqno(20)$$ In terms 
of the formfactor (20) the matrix element of the Heisenberg kink 
field can be written as $$ \<\chi| \f(x,t)|\chi'\>=\exp \left\{ 
{i \over \h} M \left[ t(\cosh \chi-\cosh \chi')-x(\sinh \chi- 
\sinh \chi')\right] \right\} \tilde f(\chi-\chi').  \eqno(21)$$ 
Taking the matrix element $\<\chi| \dots |\chi'\>$ from the eq. 
of motion (8) and using eq.(21), we find in the leading 
approximation (neglecting mesons) $$ {2M^2 \over \h^2} \left[ 
\cosh(\chi-\chi')-1\right] 
\<\chi|\f(x,t)|\chi'\>+\<\chi|V'(\f(x,t)|\chi'\>=0, \eqno(22)$$ 
whence by removing the $xt$-dependence and going over to $f(\x)$ 
by means of eq.(20) we'll get nothing else but the 
finite-difference eq.(14), provided $f(z)$ is analytic in the 
strip $| Im \ z| < \h$. The last requirement is crucial for the 
equivalence between eqs.(14) and (22), since otherwise the 
Fouriet-transform of the kinetic term in eq.(22) doesn't exist 
due to the additional exponential factor $\ch (\chi-\chi')$.

The same picture we'll find by evaluating the kink energy in the 
GJ-approach. Due to the explicit covariance it suffers to 
calculate only the kink mass for the states at rest, which is 
given by the normalized matrix element $$ M[f]= {\<\c'|H(\f)|\c 
\> \over \< \c'| \c \> } {\; \vrule height 3.5 ex depth 3 ex 
\;}_{\c=0} \ , \eqno(23)$$ where $H(\f)$ is the field 
Hamiltonian $$ H(\f)=\int \! dx \ \[ \1/2(\pd_t \f)^2+\1/2 
(\pd_x \f)^2+V(\f) \].  \eqno(24)$$ Removing the spatial 
integration, we get $$M=2\pi\h \ \d(M\sh\c-M\sh\c') \ {\< 
\c'|\ham(0)|\c \> \over \< \c'|\c \> }{\; \vrule height 3.5 ex 
depth 3 ex \;}_{\c=0}= {2\p\h \over M \ch \c } \ \< 
\c|\ham(0)|\c \>{\; \vrule height 3.5 ex depth 3 ex \;}_{\c=0} \ 
\ , \eqno(25)$$ where $\ham(x)$ is the Hamiltonian density. In 
the last expression the main attention should be paid to the 
kinetic term with the field derivatives $$ T=\1/2 \[ (\pd_t 
\f)^2+(\pd_x \f)^2 \]. \eqno(26)$$ Following GJ, by evaluating 
$\<\c|T|\c\>$ we insert into the matrix element the intermediate 
set of the kink states $ \int \!  d\chi' \ |\chi'\> \<\chi'|$ 
and find $$\<\chi|T|\chi\>={M^2 \over 2\h^2} \int \!\! \ d\chi' 
\left[ (\cosh \chi-\cosh \chi')^2+(\sinh \chi-\sinh \chi')^2 
\right] \ |\tilde f(\chi-\chi')|^2.  \eqno(27)$$ Now, if we put 
$\c=0$ and replace formally $\tilde f(\chi)$ by $ f(\xi)$, we'll 
get the following expression $$ \<\c|T|\c\>{\; \vrule height 2.5 
ex depth 2 ex \;}_{\c=0}=(2\p\h)^{-1} {M^2 \over 2\h^2} \int 
\!\! d\x \ \left\{ \[ \( \cos \h \pd_\xi -1 \) f (\x) \]^2+ \[ 
\( \sin \h \pd_\x \) f (\x) \]^2 \right\}. \eqno(28)$$ However, 
in general such transition from eq.(27) to (28) is incorrect, 
since due to the exponential factor  $\[ (\ch \c-\ch \c')^2+(\sh 
\c-\sh \c')^2 \] $ the expression (27) diverges for such 
$f(\x)$, for which (28) remains well-defined. In particular, it 
takes place when $\tilde f(\c)$ decreases for $|\c| \to \infty $ 
slowlier than $\exp(-|\c|)$, what is just the case when the pole 
of $f(z)$ appears in the strip $|Im\ z | < \h$.  It is indeed 
this circumstance, that makes the interchange of integrations by 
calculation of the matrix element $\<\chi|T|\chi\>$ in general 
illegal and so yields an infinite difference between eqs.(27) 
and (28).  This difference is nothing else, but the quantum 
descendant of the classical singularity, caused by the jump in 
the field configuration, since in the quasiclassical limit $\h 
\to 0$ the existence of the pole in the strip $|Im \ z|<\h$ 
implies, that it approaches the real axis, creating the 
discontinuity in $f(\x)$.  At the same time, the covariant 
approach to kink quantization, described above, permits us to 
separate by means of the eq.(19) the corresponding singularities 
in the field derivatives  as the additional pole terms, 
containing $\d$-functions of the complex argument.  Now let us 
present some arguments (without any claim of generality), that 
by quantization the field in the rapidity (or $\x\c$-) 
representation instead of the conventional $xp$-one, such 
$\d$-functions turn out to be the natural form of representing 
the UV-singularities of the theory. For convenience we put the 
mass of the elementary quantum of the theory equal to unity 
$m=1$, but retain the $\h$-dependence. 

In the standard $xp$-representation the canonical field 
variables are $$ \f (x)={1 \over \sqrt {2 \pi \h} } \ \int \! 
{dp \over \sqrt{2\w_p}} \ \[ a_p e^{ipx/\h} + h.c. \] \ , \quad 
\pi (x)={i \over \sqrt {2 \pi \h} } \ \int \! \sqrt{{\w_p \over 
2}} dp  \ \[ a^+_p e^{-ipx/\h} - h.c. \] \ , \eqno(29)$$ where 
$$ [a_p , a^+_{p'}]=\h \d_{pp'} \ , \quad [a_p , a_{p'}]=0. 
\eqno(30)$$ The rapidity representation $p=\sh \chi, \ \w_p=\ch 
\chi $ is introduced by the following canonical change of 
creation-annihilation operators $$ a_p=a_{\chi}/\sqrt{\ch \chi}, 
\eqno(31)$$ provided $$ [a_{\chi} , a^+_{\chi'}]=\h \d_{\chi 
\chi'} \ , \quad [a_{\chi} , a_{\chi'}]=0. \eqno(32)$$ The 
appropriate field variables for $\x\c$-representation are $$ \F 
(\x)= {1 \over \sqrt {2 \pi \h} } \ \int \!  {d\chi \over 
\sqrt{2}} \ \[ a_{\chi} e^{i\chi \x/\h} + h.c. \] \ , \quad \Pi 
(\x)={i \over \sqrt {2 \pi \h} } \ \int \!  \ch \chi \ {d\chi 
\over \sqrt{2}} \ \[ a^+_{\chi} e^{-i\chi\x /\h} - h.c.  \] \ , 
\eqno(33)$$ and obey the following  commutation relations $$ [\F 
(\x), \Pi (\x')]= {i \over 2 \pi} \ \int \!  d\chi \ \ch \chi \ 
e^{i\chi (\x-\x') /\h}= {i\h \over 2} \[ 
\d(\x-\x'+i\h)+\d(\x-\x'-i\h)\].  \eqno(34)$$ So the rapidity 
repr. contains from the beginning the $\d$-functions of the 
complex argument, quite similar to those of eq.(19). On the 
other hand, it is indeed this singularity in the field 
commutator (34), which causes ultimately all the UV-divergencies 
of the quantum theory. So it should be clear, that in the 
rapidity representation all the divergent quantities can be 
written in the same fashion. For instance, for the vacuum energy 
of the free field one has $$ E_{vac}= {1 \over 2} \ \int \! \w_p 
dp \ (a_p a^+_p - a^+_p a_p)={1 \over 2}  \ \int \!  \ch \chi \ 
d\chi \ (a_{\chi} a^+_{\chi} -a^+_{\chi}a_{\chi} ), \eqno(35)$$ 
what can be expressed through the field commutator as $$ 
E_{vac}= {1 \over 2i}  \ \int \! d\x \ d\x' \ \d(\x-\x') \ 
[\F(\x),\Pi (\x')]= {\h \over 2} \ \int \!  d\x \ d\x' \ 
\d(\x-\x') \ \1/2 [\d(\x-\x'+i\h)+\d(\x-\x'-i\h) ].  \eqno(36)$$ 
Of course, the eq.(36) as well as (35) give the same final 
divergent answer $ E_{vac} = (\h/2) \int \!  \ch \chi \ d\chi $.  
However, the main result of this exercise is that the rapidity 
repr. shows up a direct relationship between UV-divergencies and 
$\d$-functions of the complex argument. 

Motivated by these arguments, we assume now, that such 
$\d$-functions should be subtracted whenever they appear in 
calculations. For the perturbation theory this assumption leads 
to the conventional renormalization techniques, whereas in the 
present case it implies the subtraction of the pole terms from 
eq.(19). Such procedure regularizes the field derivatives for 
finite jumps and so is equivalent to an effective UV-cutoff, 
implemented in a nonperturbative way. Note, that this 
subtraction cannot be understood as a result of partial 
resummation of the regularized perturbation series, since here 
we are strictly related to the knowledge of the poles and 
residues $a_n$ and $\G_n$ of the given solution $f(\x)$, and so 
are not able (or al least do not know how) to reformulate such 
procedure as adding a finite number of universal counterterms to 
the initial Lagrangean, what is the charge for 
nonperturbativity. Moreover, even for a renormalizable theory 
the perturbation series might be divergent.  So our approach 
should be merely understood as an attempt to give sense to the 
theory in the case, when the conventional methods of 
UV-regularization do not work.

Removing the pole terms from field derivatives, we get the 
finite-difference eq.(14) as the dynamical eq. for the kink 
shape $f(\x)$.  However, now $f(z)$ is allowed to possess poles 
in the strip $| Im \ z|<\h$, it should be regular only on the 
lines $Im \ z= 0, \pm \h$.  Proceeding further, we'll find, that 
the equation for the kink mass will also contain the same 
finite-difference structure.  Namely, expressing the field 
derivatives as commutators and subtracting the pole terms, we 
represent the kinetic energy as $$ T=\1/2\[ (\partial_t 
\f)^2+(\partial_x \f)^2 \]= {i^2 \over 4\h^2} \sum \limits 
_{\pm} \ \{K_{\pm}[f(\xi)-f(\xi \mp i\h)] \ [f(\xi \pm 
i\h)-f(\xi)]K_{\pm}\}=$$ $$={i^2 \over 2\h^2} 
\{H[f(\xi)-f(\xi-i\h)][f(\xi+i\h)-f(\xi)]H \ +  \ 
P[f(\xi)-f(\xi-i\h)][f(\xi+i\h)-f(\xi)]P\}, \eqno(37)$$ with the 
same minimal conditions on $f(\x)$. The kink mass is defined as 
the matrix element (23), and to calculate $\< \c|\ham(0)|\c \>$ 
we insert the complete set $\int \!  d\x \ |\x \> \< \x | $.  
After some simple algebra the final answer is $$ Œ^2={Œ^2 \over 
2\h^2} \int \!\!  d\x \ [ f(\x+i\h)-f(\x)][f(\x-i\h)-f(\x)]+ 
\int \!\!  d\x \ V(f(\x)) \ .  \eqno(38)$$ The kinetic term here 
coincides completely with the "regularized" version (28) of $ 
\<\c|T|\c\>$, computed by the GJ-approach.

Now let us turn to the kink content of the theory after such 
regularization of finite jumps. First of all, the smooth 
solutions of the theory remain unchanged, since by definition 
they should be analytic in the strip $|Im \ z| <\h|$, so there 
are no subtractions at all. The most convenient way to see this 
is to expand the eq.(14) in powers of $\h$.  Then to the leading 
order we get $$ -M_0^2 f''(\x)+ V'(f)=0, \eqno(39) $$ where 
$M_0$ is the quasiclassical $(\h \to 0)$ limit of the quantum 
kink mass $M$. The eq.(39) is nothing else, but the classical 
stationary kink equation of motion,  provided $x=\x/M_0$. So if 
the proper limit $\h\to 0$ for $f(\x)$ exists, it should be 
equal to $$ f(\x)=\f_c(\x/M_0), \eqno(40)$$ where $\f_c (x)$ is 
the classical kink solution. Performing a similar expansion for 
the mass eq.(38), we'll find $$ M_0^2=\int \!\!  d\x \ \[ {M_0^2 
\over 2} f'(\x)^2+V(f(\x)) \], \eqno(41)$$ whence by means of 
eq.(40) we immediately obtain for $M_0$ the classical expression 
for the kink mass $$M_0= \int \!\!  dx \ \[ \1/2 \f_c'(x)^2 + 
V(\f_c(x))\].  \eqno(42) $$ However, in contrast to the 
classical eq.(39), the finite-difference eq.(14) admits for a 
set of other kink solutions, which are in some sense generated 
by the classical one, but are irregular for $\h \to 0$, and so 
should be called its "quantum copies". The derivation of such 
solutions has been discussed in detail in ref. [4]. The result 
is $$ f_N(\x)=\f_c(\a_N\x), \eqno(43)$$ where $N$ is an integer, 
$$ \a_N={1 \over M_N} + {2 \p N \over m\h}, \eqno(44) $$ with 
$M_N=M[f_N]$ being the corresponding mass, while $m$ is the mass 
of the elementary quantum (meson). The solutions (43) are exact 
up to $O(\h)$, the first correction to them appears at the 
one-loop level and so can be freely ignored to the leading 
order. For $N=0$ the eq.(43) coincides with (40), and this is 
the only case, when the solution is regular for $\h \to 0$.  For 
$N \not = 0$ we get for $M_N$ $$M_N=\sqrt {M_0 \over |\a_N|} 
\simeq \sqrt{ \h \over |N|} \( {M_0m \over 2 \p} \)^{1/2}. 
\eqno(45)$$ Therefore the masses of quantum kink copies are 
proportional to $\sqrt {\h}$ and so lye in between the classical 
kink mass, which is $O(1)$, and meson energies, whose magnitude 
is $O(\h)$.  Moreover, $M_N$ decreases for increasing $|N|$.  A 
similar behaviour reveals the size of solutions (43), which 
turns out to be $O(\h/|N|)$. So for $\h/N \to 0$ the size of the 
copy vanishes, and the copy shrinks into an almost discontinuous 
steplike configuration $f_{\infty}(\x)$ of the same topology, as 
the initial classical kink. In other words, we rediscover the 
discontinuous steplike classical solution $\f_d(x)$, introduced 
at the beginning, as the (improper) limit of quantum copies for 
$\h/N \to 0$ $$ f_N(\x) \to f_{\infty}(\x) =\f_d(\x). 
\eqno(46)$$ According to eq.(45) the mass of $f_{\infty}(\x)$  
vanishes.  This result is a direct consequence of that through 
our subtracting procedure we have introduced an effective 
UV-cutoff, because when the size of the copy decreases, the 
relative weight of the high-frequency component of the field 
increases and so yields the main (divergent) contribution to all 
the field observables, what is exactly cancelled by the proposed 
subtraction.  Mathematically this result is trivial, since 
$f_{\infty}(\x)$ is purely discontinuous, and so its derivatives 
contain only $\d$-functions, after subtraction of which the 
energy vanishes exactly. At the same time, without subtraction 
of the pole terms none of copies is possible, at least for 
1+1-dimensional kink models. This happens because in this case 
$\f_c(x)$ can be represented as a (Dirichlet) series in powers 
of $e^{\pm m x}$ (depending on the sign of $x$), so is $2\pi i 
/m$ periodic. On the other hand, for the Nth copy the strip $|Im 
\ z|<\h$ maps into $|Im \ z| < 2 \pi N /m$ for $\f_c(z)$, where 
it must be analytic for the given copy to exist without 
subtraction. So $\f_c(z)$ must be an entire function, but this 
is not the case for the kink models of interest.

  It should be noted, however, that the relation (46), being 
quite rigorous in the mathematical sense, cannot take place 
physically. First of all, for vanishing energy the NW c.m.s.  
coordinate (2) as well as the substitution (4) become 
ill-defined.  Secondly, in our approach the appearance of the 
kink (or its copy) is considered as the main dynamical effect, 
and quantum fluctuations of the field are treated only 
perturbatively, what might be correct only until the mass of the 
kink is sufficiently larger than the mesonic energies $\h \w_k$. 
Otherwise one has to consider the mesonic excitations at the 
same footing with the kink, what might  significantly disturb 
the whole dynamical picture. From this point of view the masses 
of copies should be bounded from below by $O(\h)$, and so there 
appears an upper bound $|N|_{max} \simeq O(1/\h)$, what 
corresponds to the copy with the minimal mass and size, allowed 
for given $\h$. With disregard to quantum corrections the latter 
looks like the limiting $f_{\infty}(\x)$, but the abovementioned 
difference between them is principal, since it shows that the 
formal appearance of $f_{\infty} (\x)$ with zero mass doesn't 
actually contradict the  uniqueness of the vacuum state. 

Let us also mention, that the appearance of the kink copies  is 
a highly relativisitic effect. If we restore the speed of light 
explicitly, we'll find $$ [L,P]= i\h {H \over c^2} \ , 
\eqno(47)$$ what scales the step in finite differences up to 
$i\h/c$.  Therefore there are in fact two independent parameters 
in our approach, the first one is $\h '=\h/c$ and describes the 
finite-difference effects, while the second --- $\sqrt \h$ --- 
serves as the expansion parameter for dealing with mesonic 
fluctuations. 

 To be more concrete, let us consider as an illustrative example 
the standard $\f^4$-model, when the selfinteraction of the field 
is $$ V(\f)=\1/2 (1-\f^2)^2, \eqno(48)$$ and yields the 
classical kink solution $$ \f_c(x)=\tanh x.  \eqno(49)$$ Its 
copies are $$ f_N(\x)=\tanh \a_N\x, \eqno(50)$$ where 
$\a_N=1/M_N + \p N/\h$, since now $m=2$. The functions (50) are 
meromorphic with simple poles at $a_n=i\p(n+1/2)/\a_N$, so there 
are exactly $2N$ poles in the strip $|Im\ z|<\h$. The kink 
copies  (50) do not be the exact solutions of eq.(14), but 
 satisfy the eq.  $$ \dalam f_N+ 2 \ {\tan^2 \b \over \b^2} \ 
{f_N \ (f_N^2 -1) \over 1+ f_N^2 \ \tan^2 \b }=0 \ , \eqno(51)$$ 
where $\b=\h/M_N$ and so is $O(\sqrt{\h})$. Because  
$|f_N(\x)|\leq 1$  for all real  $\x$, the difference between 
the "potential" term in eq.(50) and the correct one $2 f_N 
(f_N^2-1)$ turns out to be $O(\h)$ on the whole $\x$-axis, as it 
was stated above.

The topology of kinks (50) is the same as of the classical one, 
so they carry the same topological charge $Q=\pm 1$ and shrink 
for $\h/N \to 0$ into the discontinuous sign solution with 
vanishing mass. For the kinks (50) the last effect can be made 
even more apparent, since for $\h/N$ sufficiently small they can 
be approximated by $i\h$-periodic functions $\tanh \( \pi N\x 
 /\h \)$, obeying $f(\x \pm i\h)=f(\x)$. For such functions the 
kinetic term in the mass eq.(38) will vanish identically, while 
the potential term is nonzero only in the transition region, 
where $|f(\x)| \not =1$, and so vanishes for $\h/N \to 0 $. 

To conclude we have shown, that the finite jumps of the field 
profile might be regularized in a way, being equivalent to 
introducing an effective UV-cutoff and respecting the 
Lorentz-covariance from the very beginning.  As a result, the 
energy (mass) of the discontinuous steplike solution, defined as 
the (improper) limit of copies (43), becomes vanishing, rather 
than diverging. A number of accompanying effects, the most 
important of which is the possibility of the kink collapse into 
its copy, is discussed elsewhere [18]. 

This work has been done by partial financial support of RFFI, 
grant No.96-02-18097, S-P Concurrency Center of Fundamental 
Research, grant No. 95-0-6.3-20 and the Scientific Program 
"Russian Universities".

\vskip 5 true mm 

\end{document}